\documentclass[12pt]{article}

\usepackage[a4paper,text={16.8cm,22.4cm}]{geometry}
\usepackage{amsmath,amsfonts,braket,slashed,amssymb,tikz,bm,psfrag,graphicx,color,dsfont}
\usepackage{multicol}

\RequirePackage[sort&compress,square,comma,numbers]{natbib}
\allowdisplaybreaks
\begin{document}

\begin{titlepage}

\begin{flushright}
\normalsize
MITP/15-117\\ 
December 30, 2015
% arXiv:1512.nnnnn
% v1: December 30, 2015
\end{flushright}

\vspace{1.0cm}
\begin{center}
\Large\bf\boldmath
Exclusive Radiative $Z$-Boson Decays to Mesons\\ 
with Flavor-Singlet Components
\end{center}

\vspace{0.5cm}
\begin{center}
Stefan Alte$^a$, Matthias K\"onig$^a$ and Matthias Neubert$^{a,b}$\\
\vspace{0.7cm} 
{\sl ${}^a$PRISMA Cluster of Excellence \& Mainz Institute for Theoretical Physics\\
Johannes Gutenberg University, 55099 Mainz, Germany}\\[3mm]
${}^b$Department of Physics \& LEPP, Cornell University, Ithaca, NY 14853, U.S.A.\\
\end{center}

\vspace{0.8cm}
\begin{abstract}
We present a detailed study of the exclusive radiative decays $Z\to\eta^{(\prime)}\gamma$ employing the
QCD factorization approach. We derive a factorization formula for the decay amplitudes valid at leading power in an expansion in $(\Lambda_{\rm QCD}/m_Z)^2$, which includes convolutions of calculable hard-scattering kernels with the leading-twist quark and gluon light-cone distribution amplitudes of the mesons. Large logarithms arising in the evolution from the high scale $m_Z$ down to hadronic scales are resummed using the renormalization group, carefully accounting for the effects of the heavy bottom and charm quarks. Our results for the branching ratios are very sensitive to hadronic input parameters, such as the decay constants and mixing angle characterizing the $\eta\!-\!\eta'$ system. Using the most recent estimates of these parameters, we obtain the branching ratios $\mbox{Br}(Z\to\eta\gamma)\sim 1.6\cdot 10^{-10}$ and $\mbox{Br}(Z\to\eta'\gamma)\sim 4.7\cdot 10^{-9}$. A measurement of these processes at a future high-luminosity $Z$ factory could provide interesting information on the gluon distribution amplitude.
\end{abstract}

\end{titlepage}

\section{Introduction}

Exclusive decay processes involving individual hadrons in the final state pose a formidable challenge to theoretical physics, because the complicated strong-interaction physics describing hadronic bound states cannot be described using perturbative methods. For the case of hard exclusive processes the QCD factorization approach \cite{Lepage:1979zb,Efremov:1978rn,Chernyak:1983ej} provides a systematic framework for factorizing calculable short-distance effects associated with high energy scales from hadronic dynamics, which is described in terms of light-cone distribution amplitudes (LCDAs) of individual hadrons. This approach has been applied successfully for different processes, such as meson form factors at large momentum transfer (see \cite{Kroll:2010bf,Agaev:2010aq} for recent discussions) and hadronic weak decays of heavy $B$ mesons \cite{Beneke:1999br}. In previous work, we have studied the exclusive radiative decays of $Z\to M\gamma$ and $W\to M\gamma$ into final states containing a single meson $M$ as an ideal testing ground for the QCD factorization approach, arguing that power corrections to the factorized amplitudes are suppressed by $(\Lambda_{\rm QCD}/m_{Z,W})^2$ and are thus bound to be very small \cite{Grossmann:2015lea}. By including higher-order QCD corrections in the short-distance coefficients and solving their renormalization-group (RG) evolution equations, large logarithms of the form $\big[\alpha_s\ln(m_Z^2/\mu_0^2)\big]^n$, where $\mu_0\approx 1$~GeV is a typical hadronic scale, can be resummed to all orders of perturbation theory. Applying the same formalism to the exclusive radiative Higgs-boson decays $h\to V\gamma$, where $V$ is a vector meson, provides access to the Yukawa couplings of the Higgs boson to light quark flavors and thus serves as a powerful probe of physics beyond the Standard Model \cite{Grossmann:2015lea,Bodwin:2013gca,Kagan:2014ila,Koenig:2015pha}. 

When studying the decays $Z\to M\gamma$ in \cite{Grossmann:2015lea} one set of processes was left out, namely those where the final state pseudoscalar meson $M=P$ has a flavor-singlet component in its wave function. An important complication in this case lies in the fact that there exists a new contribution to the decay amplitude at leading order in power counting, where the meson is formed from two collinear gluons instead of a quark-antiquark pair. The existence of this contribution not only gives rise to a more complicated form of the factorization formula but also influences the RG equations satisfied by the short-distance coefficients \cite{Terentev:1980qu,Ohrndorf:1981uz,Shifman:1980dk,Baier:1981pm}. In this paper we present a detailed analysis of the decays $Z\to\eta^{(\prime)}\gamma$ in the context of QCD factorization, treating flavor mixing in the Feldmann-Kroll-Stech (FKS) scheme \cite{Feldmann:1998vh} and carefully accounting for the decoupling of the heavy bottom and charm quarks in the evolution from the high-energy scale $m_Z$ down to low energies.

\section{Theoretical Framework}

\begin{figure}
\centering
\raisebox{1ex}{\includegraphics[width=0.24\textwidth]{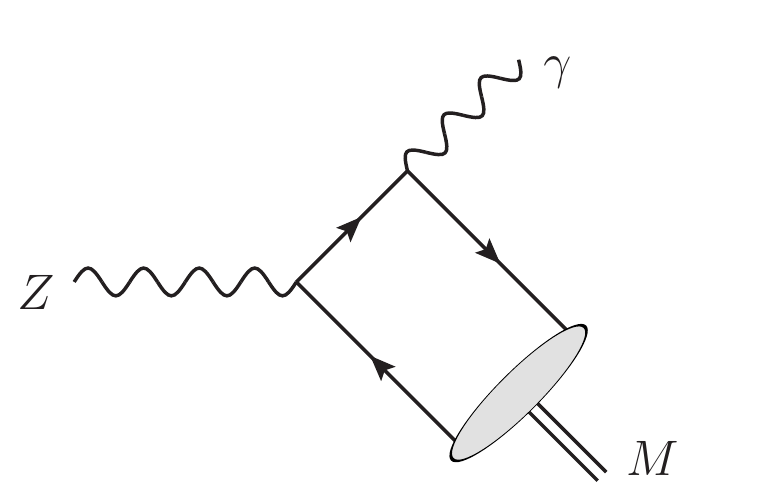}}
\includegraphics[width=0.24\textwidth]{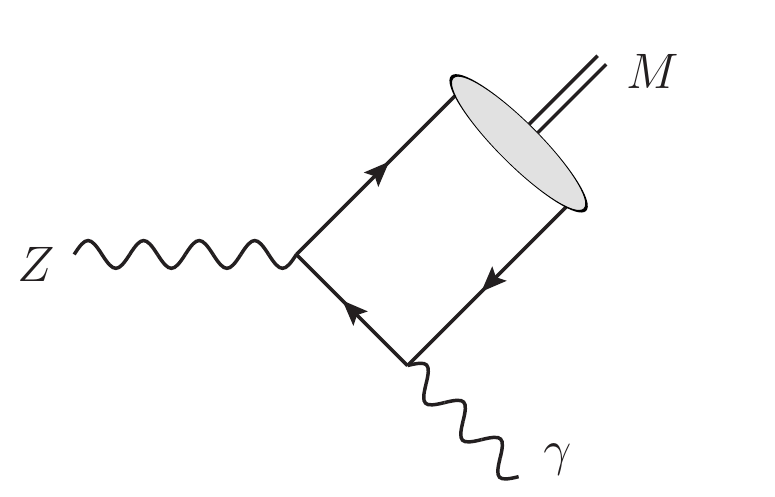}
\includegraphics[width=0.24\textwidth]{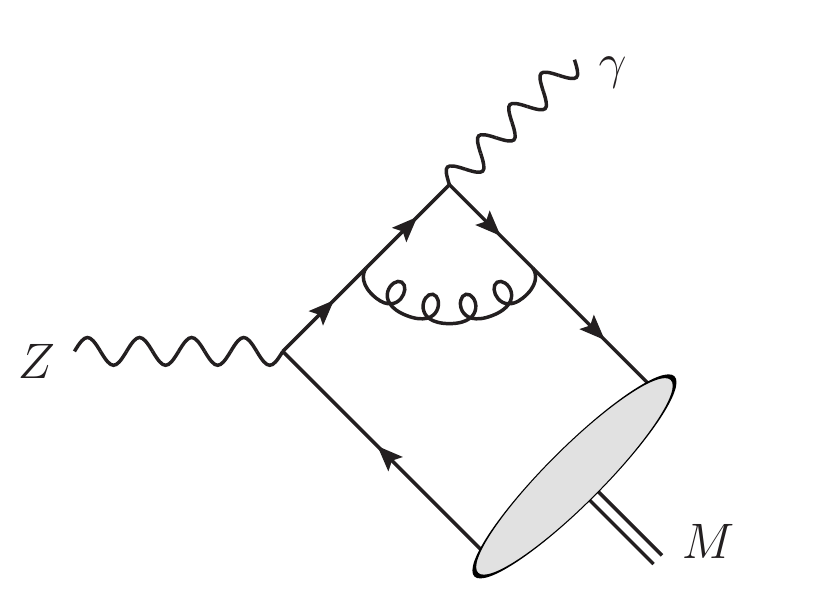}
\includegraphics[width=0.24\textwidth]{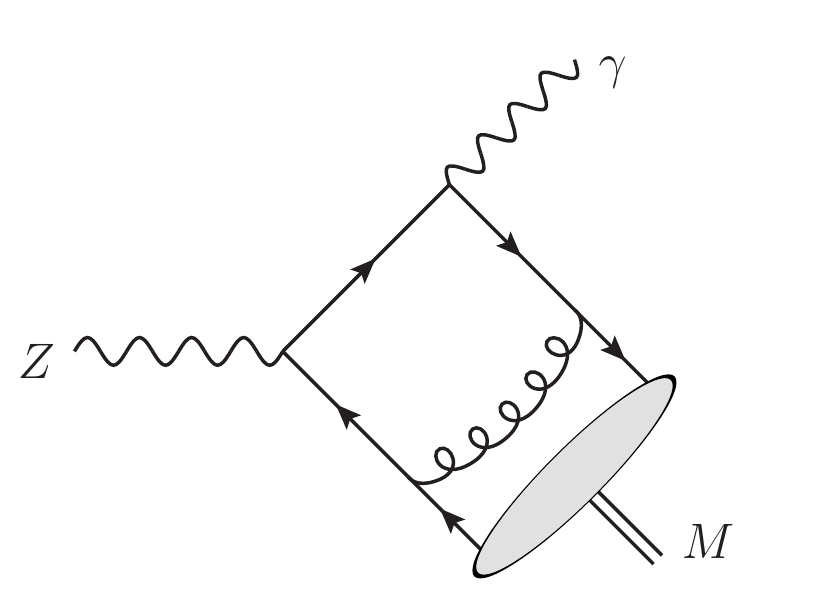}
\parbox{15.5cm}
{\caption{\label{fig:graphs} 
Some Feynman diagrams contributing to the radiative decays $Z\to M\gamma$. The meson bound state is represented by the gray blob.}}
\end{figure}

In previous work the QCD factorization formula for exclusive radiative decays $Z\to M\gamma$ was derived for the case of flavor-nonsinglet pseudoscalar or vector mesons $M=P$ or $V$, which are produced via a quark-antiquark pair \cite{Grossmann:2015lea}. Representative Feynman diagrams contributing at leading and next-to-leading order (NLO) are shown in Figure~\ref{fig:graphs}. For a pseudoscalar meson in the final state, the decay amplitude can be written in the general form 
\begin{equation}\label{ampl1}
   i{\cal A}(Z\to P\gamma)
   = \frac{eg}{2\cos\theta_W}\,i\epsilon_{\mu\nu\alpha\beta}\,
    \frac{k^\mu q^\nu\varepsilon_Z^\alpha\,\varepsilon_\gamma^{*\beta}}{k\cdot q}\,{\cal F}_P \,,
\end{equation}
where $k$ and $q$ denote the meson and photon momenta, $e$ and $g$ are the electromagnetic and weak coupling constants, and $\theta_W$ is the weak mixing angle. A second form factor, which is allowed by Lorentz invariance, vanishes since the final-state meson $P$ is an eigenstate of the charge-conjugation operator. At leading order in an expansion in powers of $(\Lambda_{\rm QCD}/m_Z)^2$, the form factor for the case of $Z\to\pi^0\gamma$ decays reads \cite{Grossmann:2015lea}
\begin{equation}\label{Fpi0}
   {\cal F}_{\pi^0} = \big( Q_u v_u - Q_d v_d \big)\,\frac{f_{\pi^0}}{\sqrt2}
    \int_0^1\!dx\,H_q(x,\mu)\,\phi_{\pi^0}(x,\mu) \,,
\end{equation}
where $f_{\pi^0}\approx 130\,\mbox{MeV}$ is the decay constant of the pion and $\phi_{\pi^0}(x,\mu)$ is its leading-twist LCDA. $Q_q$ and $v_q=\frac12\,T_3^q-Q_q\sin^2\theta_W$ are the electric and weak vector charges of a quark with flavor $q$. At NLO in QCD perturbation theory, the hard-scattering kernel $H_q(x,\mu)$ reads
\begin{equation}\label{HqNLO}
   H_q(x,\mu) = \frac{1}{x} \left[ 1 + \frac{C_F\alpha_s(\mu)}{4\pi}\,h_q(x,\mu) 
    + {\cal O}(\alpha_s^2) \right] + (x\to 1-x) \,,
\end{equation}
where $C_F=4/3$ is a color factor, and \cite{Braaten:1982yp} 
\begin{equation}
   h_q(x,\mu) = (2\ln x+3) \left( \ln\frac{m_Z^2}{\mu^2} - i\pi \right) 
    + \ln^2 x - \frac{x\ln x}{1-x} - 9 \,.
\end{equation}
The convolution integral of the hard-scattering kernel with the LCDA in (\ref{Fpi0}) is independent of the choice of the factorization scale $\mu$. The $Z\to\pi^0\gamma$ branching ratio turns out to be strongly suppressed, $\mbox{Br}(Z\to\pi^0\gamma)=(9.80\pm 1.03)\cdot 10^{-12}$ \cite{Grossmann:2015lea}, because the relevant combination of quark charges\begin{equation}
   Q_u v_u - Q_d v_d = \frac{1-4\sin^2\theta_W}{12}\approx 6.2\cdot 10^{-3}
\end{equation}
is numerically very small. We use $\sin^2\theta_W=0.23126(5)$ as extracted from the neutral-current couplings of the $Z$ boson \cite{Agashe:2014kda}. It is thus more promising to search for the decays $Z\to\eta^{(\prime)}\gamma$. Two complications arise in this case. First, the physical $\eta$ and $\eta'$ mesons are complicated mixtures of quark-antiquark states with different flavor. Second, and more profoundly, the flavor-singlet quark-antiquark state mixes with a pure gluon state under renormalization, and indeed the $\eta$ and $\eta'$ mesons can also be produced via a two-gluon LCDA.

\subsection{Factorization in the presence of flavor-singlet contributions}

The $Z\to P\gamma$ decay amplitudes, where $P=\pi^0,\eta,\eta',\dots$ denotes a neutral pseudoscalar meson, can be calculated from first principles using the QCD factorization approach \cite{Lepage:1979zb,Efremov:1978rn,Chernyak:1983ej}, because the energy $E$ released to the final-state meson is much larger than the scale of long-distance hadronic physics. At leading power in an expansion in $\lambda=\Lambda_{\rm QCD}/m_Z$, the form factors ${\cal F}_P$ can be written as convolutions of calculable hard-scattering coefficients with leading-twist LCDAs of the meson $P$. To derive the corresponding factorization formula we employ the formalism of soft-collinear effective theory (SCET) \cite{Bauer:2000yr,Beneke:2002ph}, which provides a systematic expansion of decay amplitudes in powers of $\lambda$. For the purposes of this discussion we work in the rest frame of the decaying $Z$ boson and assign momenta $k^\mu=En^\mu$ and $q^\mu=E\bar n^\mu$ to the meson and photon, respectively, where $E=m_Z/2$, while $n^\mu=(1,0,0,1)$ and $\bar n^\mu=(1,0,0,-1)$ are two light-like vectors. Up to power corrections of order $(m_P/m_Z)^2$ the meson mass can be set to zero. The light final-state meson moving along the direction $n^\mu$ can be described in terms of collinear quark, antiquark and gluon fields. These particles carry collinear momenta $p_c$ that are approximately aligned with the direction $n$. Their components scale like $(n\cdot p_c, \bar n\cdot p_c, p_c^\perp)\sim E(\lambda^2,1,\lambda)$. Note that $p_c^2\sim\Lambda_{\rm QCD}^2$, as appropriate for an exclusive hadronic state. The collinear quark and gluon fields are introduced as gauge-invariant objects dressed with Wilson lines. Explicitly, one defines \cite{Bauer:2001ct,Hill:2002vw}
\begin{equation}\label{SCETfields}
   {\cal X}_c = \frac{\rlap{\hspace{0.02cm}/}{n}\rlap{\hspace{0.02cm}/}{\bar n}}{4}\,
    W_c^\dagger\,q \,, \qquad
   {\cal A}_{c\perp}^\mu = W_c^\dagger\,(iD_{c\perp}^\mu W_c) \,, 
\end{equation}
where $iD_c^\mu=i\partial^\mu+g_s A_c^\mu$ denotes the covariant collinear derivative, and $W_c$ is a collinear Wilson line extending from the location of the field to infinity along the direction $\bar n$. Both fields are of ${\cal O}(\lambda)$ in SCET power counting, while other components of the gluon field are of higher order. Adding more component fields to an operator thus always leads to further power suppression in $\lambda$. At leading order, the operators with a non-zero matrix element between the vacuum and a single meson state are thus bilinears of the form ${\bar{\cal X}}_c\,\dots\,{\cal X}_c$ and ${\cal A}_{c\perp}^\mu\,\dots\,{\cal A}_{c\perp}^\nu$. 

Since the effective collinear fields are gauge invariant by themselves, composite operators built out of these fields can be non-local along the light-like direction $\bar n$ without leading to any power suppression \cite{Bauer:2000yr,Beneke:2002ph,Bauer:2001ct,Hill:2002vw}. The matrix elements of the bilocal quark-antiquark operators between a meson state and the vacuum can be parameterized in terms of the leading-twist quark LCDA. Specifically, one defines the {\em flavor-specific\/} quark-antiquark LCDAs
\begin{equation}\label{quarkLCDA}
\begin{aligned}
   \langle P(k)|\,\bar{\cal X}_c(t\bar n)\,\rlap{\hspace{0.02cm}/}{\bar n}\gamma_5\,{\cal X}_c(0) |0\rangle
   &= \langle P(k)|\,\bar q(t\bar n)\,\rlap{\hspace{0.02cm}/}{\bar n}\gamma_5\,[t\bar n,0]\,q(0) |0\rangle \\
   &= - i\bar n\cdot k\,f_P^q(\mu) \int_0^1\!dx\,e^{ixt\bar n\cdot k}\,\phi_P^q(x,\mu) \,,
\end{aligned}
\end{equation}
where $[t\bar n,0]=W_c(t\bar n)\,W_c^\dagger(0)$ is a Wilson line extending from 0 to the point $t\bar n$, and $q$ can be any quark flavor. The quark LCDAs are normalized such that $\int_0^1 dx\,\phi_P^q(x,\mu)=1$. The flavor-specific decay constants $f_P^q$ entering in (\ref{quarkLCDA}) are defined in terms of the local matrix elements 
\begin{equation}\label{fPqdef}
   \langle P(k)|\,\bar q\,\gamma^\mu\gamma_5 q\,|0\rangle = -if_P^q(\mu)\,k^\mu \,.
\end{equation}
Note that, due to the axial anomaly, the flavor-diagonal axial currents are not conserved. As a consequence the decay constants $f_P^q(\mu)$ are scale-dependent quantities with an anomalous dimension that starts at two-loop order \cite{Kodaira:1979pa}. 

For the matrix element of the bilocal two-gluon current we define
\begin{equation}\label{gluonLCDA}
   \frac{1}{g_s^2}\,\langle P(k)|\,\mbox{tr}\big[ {\cal A}_{c\perp}^\mu(t\bar{n})\,
    \epsilon_{\mu\nu}^\perp\,{\cal A}_{c\perp}^\nu(0) \big]\,|0\rangle
   = T_F\,f_P^{uds}(\mu) \int_0^1 dx\,\frac{e^{ixt\bar n\cdot k}}{x(1-x)}\,\phi_P^g(x,\mu) \,,
\end{equation}
where $T_F=1/2$, $\epsilon_{\mu\nu}^\perp\,(n\cdot\bar n)=\epsilon_{\mu\nu\alpha\beta}\,\bar n^\alpha n^\beta$, and we use the convention that $\epsilon^{0123}=+1$. The trace in this expression acts in color space. The normalization to the flavor-singlet sum of the light-flavor decay constants, $f_P^{uds}\equiv f_P^u+f_P^d+f_P^s$, is chosen for convenience. $C$-parity requires that $\phi_P^g(x,\mu)$ is odd under $x\leftrightarrow(1-x)$ and hence its normalization integral vanishes. Using the integral representation \cite{Hill:2002vw}
\begin{equation}
   {\cal A}_c^\mu(z) = \int_{-\infty}^0\!ds\,\bar n_\alpha
    \big[ W_c^\dagger\,g_s G_c^{\alpha\mu}\,W_c \big](z+s\bar n) \,,
\end{equation}
it is straightforward to show that (\ref{gluonLCDA}) is equivalent to the more conventional definition \cite{Beneke:2002jn}\footnote{Our distribution amplitude $\phi_P^g$ is equal to $C_F/6$ times the function $\phi_M^{(g)}$ defined in (18) of \cite{Agaev:2014wna}, $-\sqrt{C_F/12}$ times the function $\phi_{Pg}$ defined in (A.10) of \cite{Kroll:2002nt}, and $-C_F/6$ times the LCDA assumed in Sections~3 and~4 of \cite{Kroll:2002nt} and in \cite{Kroll:2013iwa}.}
\begin{equation}
   \langle P(k)|\,\bar n_\alpha\bar n_\beta\,G_{~\mu,A}^\alpha(t\bar{n})\,[t\bar n,0]_{AB}\,
    \tilde G_B^{\beta\mu}(0) |0\rangle 
   = (\bar n\cdot k)^2\,f_P^{uds}(\mu) \int_0^1\!dx\,e^{itx\bar n\cdot k}\,\phi_P^g(x,\mu) \,,
\end{equation}
where $\tilde G^{\mu\nu}=\frac12\epsilon^{\mu\nu\alpha\beta}\,G_{\alpha\beta}$ is the dual field-strength tensor, and $[t\bar n,0]_{AB}$ denotes a Wilson line in the adjoint representation of the gauge group. The gauge-invariant bilocal matrix elements in (\ref{quarkLCDA}) and (\ref{gluonLCDA}) can be multiplied by functions of the coordinate $t$. After Fourier transformation to momentum space, these function become the hard-scattering kernels.

\begin{figure}
\centering
\includegraphics[width=0.33\textwidth]{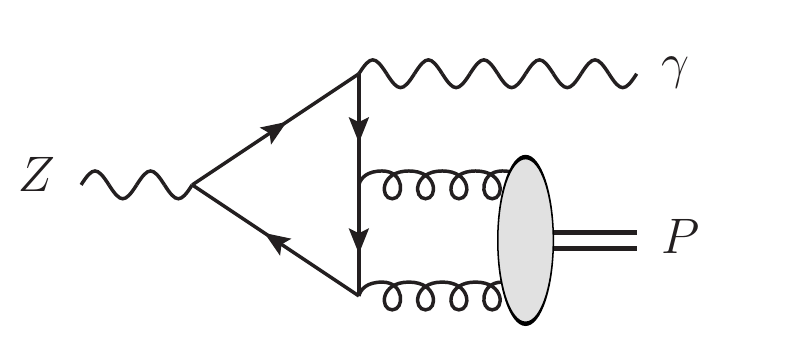}
\includegraphics[width=0.33\textwidth]{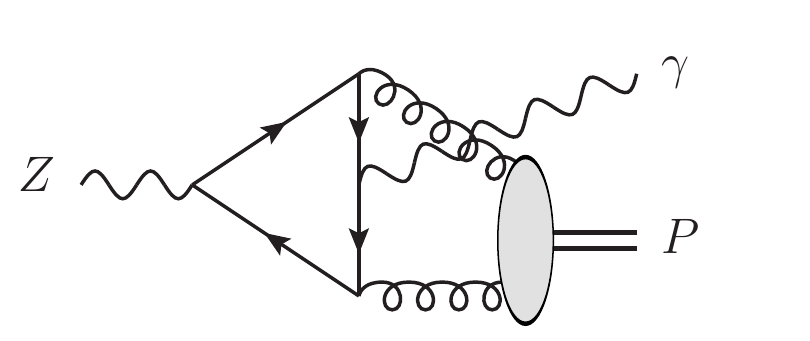}
\parbox{15.5cm}
{\caption{\label{fig:zggg} 
Representative one-loop diagrams contributing to the $Z\to\eta^{(\prime)}\gamma$ decay amplitudes, in which the final-state meson is produced via the leading-twist gluon LCDA. Other four diagrams arise from permutations of the gluon and photon legs.}}
\end{figure}

The diagrams in Figure~\ref{fig:graphs} produce all $n_f$ active quark flavors with an amplitude proportional to $Q_q v_q$. This feature remains true when QCD corrections are included. We can decompose the result into a flavor-singlet and a flavor-nonsinglet contribution. In addition, at NLO in $\alpha_s$ there exist diagrams of the form shown in Figure~\ref{fig:zggg}, in which the meson $P$ is produced via a two-gluon state. In these graphs all possible quark flavors including the top quark contribute in the loop. At a high matching scale $\mu\sim m_Z$ the heavy-particle scales $m_t$ and $m_Z$ are integrated out and absorbed into short-distance coefficient functions. Adding up the various contributions, we obtain the factorization formula 

\begin{equation}\label{Feta}
\begin{aligned}
   {\cal F}_P &= {\cal Q}_S^{(5)}\,\bigg[ \int_0^1\!dx\,H_q^S(x,\mu) \sum_q f_P^q(\mu)\,\phi_P^q(x,\mu)
    + f_P^{uds}(\mu) \int_0^1\!dx\,H_g(x,\mu)\,\phi_P^g(x,\mu) \bigg] \\
   &\quad\mbox{}+ \big( Q_u v_u - Q_d v_d \big)
    \int_0^1\!dx\,H_q(x,\mu)\,\sum_q c_q^{(5)} f_P^q(\mu)\,\phi_P^q(x,\mu) \,, 
\end{aligned}
\end{equation}
where $n_f=5$ is the number of active quark flavors in the effective theory below the scale $\mu\sim m_Z$, and in general we define
\begin{equation}
   {\cal Q}_S^{(n_f)} = \frac{1}{n_f} \sum_q Q_q v_q \,,
    \quad \mbox{and} \quad
   Q_q v_q - {\cal Q}_S^{(n_f)} \equiv \big( Q_u v_u - Q_d v_d \big)\,c_q^{(n_f)} \,.
\end{equation}
The terms in the first line in (\ref{Feta}) correspond to the flavor-singlet contributions to the form factors, involving both quark and gluon LCDAs. The terms in the second line are the combined flavor-nonsinglet contributions, with $\sum_q c_q^{(n_f)}=0$. In Table~\ref{tab:coefs} we collect the relevant coefficients for different numbers of active flavors. We will discuss in Section~\ref{sec:RGevol} how (\ref{Feta}) is evolved down to a low value $\mu_0=1$\,GeV of the factorization scale. In this process the heavy bottom and charm quarks are integrated out, and the factorization formula must be matched onto an analogous formula in a low-energy effective theories with $n_f=3$ active flavors.

In order to compute the hard-scattering kernels in (\ref{Feta}) at NLO in perturbation theory we evaluate the diagrams shown in Figure~\ref{fig:graphs} (plus six other one-loop graphs) for $H_q^{(S)}(x,\mu)$, and those shown in Figure~\ref{fig:zggg} for $H_g(x,\mu)$, using dimensional regularization with $d=4-2\epsilon$ space-time dimensions and working in the $\overline{\rm MS}$ scheme. In practice, one evaluates these diagrams with on-shell external quark and gluon states carrying momenta $xk^\mu$ and $(1-x) k^\mu$, and then applies projections onto the meson LCDAs. The relevant projections for the quark LCDAs have been discussed in \cite{Grossmann:2015lea}. For the gluon case one computes the partonic amplitude in the form ${\cal A}_{\rm part}=A_{\mu\nu}(x,\mu)\,\varepsilon_A^{*\mu}(xk)\,\varepsilon_B^{*\nu}((1-x)k)$, where $A,B$ are color indices. The corresponding hadronic amplitude is then obtained by stripping off the gluon polarization vectors and contracting the indices with the projector \cite{Beneke:2002jn}
\begin{equation}\label{eq:gluonProjector}
   M^{\mu\nu}(x,\mu)
   = \frac{\delta_{AB}}{N_c^2-1}\,\frac{\epsilon_\perp^{\mu\nu}}{2}\,f_P^{uds}(\mu)\,
    \frac{\phi_P^g(x,\mu)}{x(1-x)} \,.
\end{equation}
The individual loop graphs contain divergences. The ultraviolet divergences cancel in the sum of all diagrams, while infrared divergences are subtracted when we renormalize the meson LCDAs (including a finite renormalization of the axial current). For the flavor-nonsinglet case this has been discussed in detail in \cite{Grossmann:2015lea}. In the flavor-singlet case one needs to account for the mixing between quark and gluon LCDAs. Organizing the products $\sum_q f_P^q\,\phi_P^q(x)$ and $f_P^{uds}\,\phi_P^g(x)$ into a two-component vector $\overrightarrow{f\phi}_P(x)$, we can express the bare functions in terms of renormalized functions via
\begin{equation}\label{RGEphi}
   \overrightarrow{f\phi}_P^\mathrm{bare}(x) 
   = \int_0^1\!dy\,\bm{Z}_{f\phi}^{-1}(x,y,\mu)\,\overrightarrow{f\phi}_P(y,\mu) \,,
\end{equation}
where the matrix $\bm{Z}_{f\phi}(x,y,\mu)$ of renormalization factors is given by
\begin{equation}\label{eq:LCDArenZ}
   \bm{Z}_{f\phi}(x,y,\mu) = \delta(x-y) + \frac{\alpha_s(\mu)}{4\pi\epsilon}
    \begin{pmatrix} V_{qq}(x,y) & ~V_{qg}(x,y) \\
                    V_{gq}(x,y) & ~V_{gg}(x,y) 
    \end{pmatrix} + {\cal O}(\alpha_s^2) \,.
\end{equation}
Explicit expressions for the kernel functions $V_{ij}(x,y)$ were obtained in \cite{Terentev:1980qu,Ohrndorf:1981uz,Shifman:1980dk,Baier:1981pm} and are collected in Appendix~\ref{app:RGEstuff}. From these equations the counterterms required to cancel the IR divergences in the bare hard-scattering coefficients are derived. At NLO in $\alpha_s$ (but not beyond), the flavor-singlet hard-scattering kernel $H_q^S(x,\mu)$ is given by the same expression as $H_q(x,\mu)$ in (\ref{HqNLO}). For the gluon kernel we find
\begin{equation}\label{HgNLO}
   H_g(x,\mu) = \frac{T_F\alpha_s(\mu)}{4\pi} \left[ 5 h_g(x,\mu) 
   + \frac{Q_t v_t}{{\cal Q}_S^{(5)}}\,h_g^t(x) - (x\to 1-x) \right] + {\cal O}(\alpha_s^2) \,,
\end{equation}
where the first term accounts for the contributions of the $n_f=5$ light quark flavors, while the second term describes the contribution of the heavy top quark. We obtain
\begin{equation}
\begin{aligned}
   h_g(x,\mu) &= - \frac{8\ln x}{(1-x)^2} \left[ \left( \ln\frac{m_Z^2}{\mu^2} - i\pi \right)
    + \frac{\ln x}{2} + \frac{1}{x} - 3 \,\right] , \\
   h_g^t(x) &= \left( \frac{2 r_t^3}{315} + \frac{r_t^4}{504} \right) (1-2x) + {\cal O}(r_t^5) \,,
\end{aligned}    
\end{equation}
where $r_t=m_Z^2/m_t^2$. The function $h_g$ agrees with the corresponding kernel for the $\gamma\gamma^*\to gg$ scattering amplitude calculated in \cite{Kroll:2002nt}. The function $h_g^t$ can be derived from the expression for the charm-quark contribution to the $\gamma\gamma^*\to\eta^{(\prime)}$ form factors presented in \cite{Agaev:2014wna}. We find that the top-quark contribution is extremely small and can be neglected for all practical purposes.

\begin{table}
\centering
\begin{tabular}{|c|c||cc|}
\hline 
$n_f$ & ${\cal Q}_S^{(n_f)}$ & Up-type quarks & Down-type quarks \\
\hline
5 & $\phantom{\Big|} \frac{7}{60}-\frac{11}{45}\,\sin^2\theta_W\approx 0.0601 \phantom{\Big|}$
 & $c_{u,c}^{(5)}=\frac35$ & $c_{d,s,b}^{(5)}=-\frac25$ \\[1mm] 
4 & $\frac18-\frac{5}{18}\,\sin^2\theta_W\approx 0.0608$ & $c_{u,c}^{(4)}=\frac12$
 & $c_{d,s}^{(4)}=-\frac12$ \\[1mm]
3 & $\frac19-\frac29\,\sin^2\theta_W\approx 0.0597$ & $c_u^{(3)}=\frac23$ & $c_{d,s}^{(3)}=-\frac13$ \\[1mm]
\hline 
\end{tabular}
\parbox{15.5cm}
{\caption{\label{tab:coefs} 
Flavor-number dependent coefficients entering the factorization formula (\ref{Feta}).}}
\end{table} 

It will be convenient to transform the factorization formula (\ref{Feta}) from momentum space to Gegenbauer moment space, as this turns the convolution integrals into simple sums. To this end one expands the meson LCDAs into Gegenbauer polynomials, such that\footnote{Our Gegenbauer moments $b_n^P$ are equal to $\frac29$ times the Gegenbauer moments $c_{n,M}^{(g)}$ used in \cite{Agaev:2014wna}, and $-\frac{1}{135}$ times the Gegenbauer moments $B_{Pn}^g$ and $a_{Pn}^g$ used in \cite{Kroll:2002nt} and \cite{Kroll:2013iwa}.}
\begin{equation}\label{eq:LCDAexpansion}
\begin{aligned}
   \phi_P^q(x,\mu) &= 6 x(1-x)\,\bigg[ 1
    + \sum_{n=2,4,\dots} a_n^{P,q}(\mu)\,C_n^{(3/2)}(2x-1) \bigg] \,, \\
   \phi_P^g(x,\mu) &= 30 x^2(1-x)^2 \sum_{n=2,4,\dots} b_n^P(\mu)\,C_{n-1}^{(5/2)}(2x-1) \,.
\end{aligned}
\end{equation}
$C$-parity implies that the quark LCDA is an even function under $x\leftrightarrow(1-x)$, while the gluon LCDA is odd. When the Gegenbauer expansions are used, the factorization formula (\ref{Feta}) in the 5-flavor effective theory can be recast in the form
\begin{equation}\label{Fetan}
\begin{aligned}
   {\cal F}_P &= 6 {\cal Q}_S^{(5)} \left[ \sum_n C_n^S(\mu) \sum_q f_P^q(\mu)\,a_n^{P,q}(\mu) 
    + \sum_n D_n(\mu)\,f_P^{uds}(\mu)\,b_n^P(\mu) \right] \\
   &\quad\mbox{}+ 6\,\big( Q_u v_u - Q_d v_d \big)
    \sum_n C_n(\mu) \sum_q c_q^{(5)} f_P^q(\mu)\,a_n^{P,q}(\mu) \,, 
\end{aligned}
\end{equation}
where the sums run over even integers $n=0,2,4,\dots$, and we define $a_0^{P,q}\equiv 1$ and $b_0^P\equiv 0$. $C_n^{(S)}(\mu)$ are the moment-space representations of the hard-scattering kernels $H_q^{(S)}(x,\mu)$, while $D_n(\mu)$ is the moment-space expression for the gluon kernel $H_g(x,\mu)$. Note that we have extracted an overall factor~6 for convenience. These coefficients can be calculated using a technique described in \cite{Grossmann:2015lea}. At one-loop order we obtain
\begin{equation}\label{Cndef}
   C_n^{(S)}(\mu) = 1 + \frac{C_F\alpha_s(\mu)}{4\pi}\,c_n(\mu) + {\cal O}(\alpha_s^2) \,, \qquad
   D_n(\mu) = \frac{T_F\alpha_s(\mu)}{4\pi}\,\bigg[ 5 d_n(\mu)
    + \frac{Q_t v_t}{{\cal Q}_S^{(5)}}\,d_n^t \bigg] + {\cal O}(\alpha_s^2) \,,
\end{equation}
with
\begin{equation}\label{gorgeous}
\begin{aligned}
   c_n(\mu) &= - \left[ 4 H_{n+1} - 3 - \frac{2}{(n+1)(n+2)} \right] 
    \left( \ln\frac{m_Z^2}{\mu^2} - i\pi \right) \\[-1mm]
   &\quad\mbox{}+ 4 H_{n+1}^2 - \frac{4H_{n+1}-3}{(n+1)(n+2)} + \frac{2}{(n+1)^2 (n+2)^2} - 9 \,, \\[1mm]
   d_n(\mu) &= \frac{20 n(n+3)}{3(n+1)(n+2)} 
    \left[ \left( \ln \frac{m_Z^2}{\mu^2} - i\pi\right) - 2 H_{n+1} - 1 + \frac{1}{(n+1)(n+2)} \right] , 
    \\[1mm]
   d_n^t &= - \left( \frac{2 r_t^3}{1323} + \frac{5 r_t^4}{10584} \right) \delta_{n2}
    + {\cal O}(r_t^5) \,.
\end{aligned}
\end{equation}
Here $H_{n+1}=\sum_{k=1}^{n+1} \frac{1}{k}$ are the harmonic numbers. The coefficients $C_n(\mu)$ and $C_n^S(\mu)$ start to differ from two-loop order on.

\subsection{Renormalization-group evolution and resummation}
\label{sec:RGevol}

Due to the large energy released to the final-state particles, the $Z\to P\gamma$ decay amplitudes receive contributions from quantum fluctuations with virtualities ranging from the high scale $m_Z$ down to the hadronic scale $\mu_0\sim 1$\,GeV characteristic for light mesons. Phenomenological input for the decay constants $f_P^q(\mu)$ and the meson LCDAs $\phi_P^q(x,\mu)$ and $\phi_P^g(x,\mu)$ is usually provided at such a low scale. The enormous scale hierarchy gives rise to large logarithms of the form $\big[\alpha_s\ln(m_Z^2/\mu_0^2)\big]^n$ in the expressions for the hard-scattering kernels -- see e.g.\ (\ref{gorgeous}) -- which need to be resummed to all orders of perturbation theory. A subtlety is that the appropriate effective theory at different scales $\mu$ between $m_Z$ and $\mu_0$ contains different number of active quark flavors. At the high scale $\mu_Z\sim m_Z$ the appropriate effective theory contains $n_f=5$ active flavors, which up to corrections of order $(m_q^2/m_Z^2)$ can be treated as massless. When the factorization scale is lowered below the threshold of the $b$-quark mass one must match relation (\ref{Fetan}) onto a corresponding expression in an effective theory containing $n_f=4$ light flavors. The flavor-singlet and flavor-nonsinglet contributions need to be rearranged in this step in order to account for the change in the coefficients collected in Table~\ref{tab:coefs}. At the same time, the contribution from the $b$-quark stops running below $\mu_b\sim m_b$ and hence it needs to be evaluated at that scale. A similar procedure takes place when crossing the charm threshold at a scale $\mu_c\sim m_c$. The final factorization theorem at the low scale $\mu_0\sim 1$\,GeV can thus be written in the form  
\begin{equation}\label{Fetanf3}
\begin{aligned}
   {\cal F}_P &= 6 {\cal Q}_S^{(3)} \left[ \sum_n C_n^S(\mu_0) \!\sum_{q=u,d,s}\! 
    f_P^q(\mu_0)\,a_n^{P,q}(\mu_0) 
    + \sum_n D_n(\mu_0)\,f_P^{uds}(\mu_0)\,b_n^P(\mu_0) \right] \\
   &\quad\mbox{}+ 6\,\big( Q_u v_u - Q_d v_d \big)
    \sum_n C_n(\mu_0) \!\sum_{q=u,d,s}\! c_q^{(3)} f_P^q(\mu_0)\,a_n^{P,q}(\mu_0) \\
   &\quad\mbox{}+ 6 Q_b v_b \sum_n C_n^b(\mu_b)\,f_P^b(\mu_b)\,a_n^{P,b}(\mu_b)
    + 6 Q_c v_c \sum_n C_n^c(\mu_c)\,f_P^c(\mu_c)\,a_n^{P,c}(\mu_c) \,.
\end{aligned}
\end{equation}
Numerical estimates of the hadronic input parameters -- the various decay constants and Gegenbauer moments -- will be provided in Section~\ref{sec:mesonMixing}.

The scale evolution of the hard-scattering coefficients $C_n^{(S)}(\mu)$ and $D_n(\mu)$, and of the decay constants and LCDAs, is controlled by RG evolution equations. The terms shown in the first and second line in the factorization formula (\ref{Feta}) are separately scale independent. However, in the first line there is a non-trivial mixing between the quark- and gluon-initiated contributions. Relation (\ref{RGEphi}) implies the RG evolution equation
\begin{equation}\label{RGevol1}
   \mu\,\frac{d}{d\mu}\,\overrightarrow{f\phi}_P(x,\mu) 
   = - \int_0^1\!dy\,\bm{\Gamma}(x,y,\mu)\,\overrightarrow{f\phi}_P(y,\mu) \,,
\end{equation}
where the anomalous-dimension matrix is given by 
\begin{equation}
  \bm{\Gamma}(x,y,\mu) = 2\alpha_s\,\frac{\partial}{\partial\alpha_s} \bm{Z}_{f\phi}^{[1]}(x,y,\mu)
  = \frac{\alpha_s(\mu)}{2\pi}\,
   \begin{pmatrix} V_{qq}(x,y) & ~V_{qg}(x,y) \\
                   V_{gq}(x,y) & ~V_{gg}(x,y) 
   \end{pmatrix} + {\cal O}(\alpha_s^2) \,.
\end{equation}
The quantity $\bm{Z}_{f\phi}^{[1]}$ is the coefficient of the $1/\epsilon$ pole in the matrix $\bm{Z}_{f\phi}$. The eigenfunctions of the one-loop evolution kernels collected in Appendix~\ref{app:RGEstuff} are Gegenbauer polynomials of rank 3/2 and 5/2, as used in (\ref{eq:LCDAexpansion}). At one-loop order the RG equation (\ref{RGevol1}) thus implies an evolution equation in the space of Gegenbauer moments, which is diagonal in $n$, namely
\begin{equation}\label{RGevol1n}
   \left[ \mu\,\frac{d}{d\mu} + \frac{\alpha_s(\mu)}{4\pi}\,
    \begin{pmatrix} \gamma_n^{qq} & ~\gamma_n^{qg} \\
                    \gamma_n^{gq} & ~\gamma_n^{gg} \end{pmatrix} + {\cal O}(\alpha_s^2) \right]
    \begin{pmatrix} \sum_q f_P^q(\mu)\,a_n^{P,q}(\mu) \\ 
                    f_P^{uds}(\mu)\,b_n^P(\mu) \end{pmatrix}
   = 0 \,,
\end{equation}
where \cite{Terentev:1980qu,Ohrndorf:1981uz,Shifman:1980dk,Baier:1981pm}
\begin{equation}
\begin{aligned}
   \gamma_n^{qq} &= 2 C_F\! \left[ 4 H_{n+1} -3 - \frac{2}{(n+1)(n+2)} \right] , ~~ &
    \gamma_n^{qg} &= - T_F n_f\,\frac{40n(n+3)}{3(n+1)(n+2)} \,, \\
   \gamma_n^{gq} &= - C_F\,\frac{12}{5(n+1)(n+2)} \,, &
    \gamma_n^{gg} &=  2C_A\! \left[ 4 H_{n+1} - \frac{8}{(n+1)(n+2)} \right] -2\beta_0 \,.
\end{aligned}
\end{equation}
Here $\beta_0=\frac{11}{3}\,C_A-\frac43\,T_F n_f$ with $C_A=3$ is the first coefficient of the QCD $\beta$-function. The corresponding evolution equation for flavor-nonsinglet combinations of the Gegenbauer moments of quark LCDAs is multiplicative and governed by the anomalous-dimension coefficients $\gamma_n^{qq}$. The solution of equation (\ref{RGevol1n}) is discussed in Appendix~\ref{app:RGEstuff}. One finds that both eigenvalues of the moment-space anomalous-dimension matrix are positive and grow logarithmically at large $n$, and the same holds for the coefficients $\gamma_n^{qq}$. It follows that all Gegenbauer moments vanish for asymptotically large values of the renormalization scale, $a_n^{P,q}(\mu)\to 0$ and $b_n^P(\mu)\to 0$ for $\mu\to\infty$, irrespective of their values at a low hadronic scale. In momentum space, this implies the asymptotic forms
\begin{equation}
   \lim_{\mu\to\infty}\,\phi_P^q(x,\mu) = 6x(1-x) \,, \qquad
   \lim_{\mu\to\infty}\,\phi_P^g(x,\mu) = 0 \,.
\end{equation}
The high value of the hard matching scale $\mu\sim m_Z$ in $Z\to P\gamma$ decays ensures that one is rather close to the asymptotic limit \cite{Grossmann:2015lea}, and this fact reduces the sensitivity of our predictions to the Gegenbauer moments of the LCDAs, which are currently not known with good accuracy.

RG invariance of the form factors in (\ref{Fetan}) implies that the flavor-singlet hard-scattering coefficients obey the evolution equation
\begin{equation}\label{RGevol2}
   \bigg[ \mu\,\frac{d}{d\mu} - \frac{\alpha_s(\mu)}{4\pi}\,
     \begin{pmatrix} \gamma_n^{qq} & ~\gamma_n^{qg} \\ \gamma_n^{gq} & ~\gamma_n^{gg} \end{pmatrix}^T
    + {\cal O}(\alpha_s^2) \bigg]
    \begin{pmatrix} C_n^S(\mu) \\ D_n(\mu) \end{pmatrix}
   = 0 \,. 
\end{equation}
The flavor-nonsinglet coefficients $C_n(\mu)$ obeys an analogous equation without mixing, in which the anomalous-dimension matrix is replaced by the coefficients $\gamma_n^{qq}$. The solution to these evolution equations obtained at leading order can be written in the form
\begin{equation}\label{Umatr}
   \begin{pmatrix} C_n^S(\mu_1) \\ D_n(\mu_1) \end{pmatrix}
    = \bm{U}_n^S(\mu_1,\mu_2) \begin{pmatrix} C_n^S(\mu_2) \\ D_n(\mu_2) \end{pmatrix} , \qquad
   C_n(\mu_1) = U_n(\mu_1,\mu_2)\,C_n(\mu_2) \,.
\end{equation}
Explicit expressions for the evolution functions $\bm{U}_n^S(\mu_1,\mu_2)$ and $U_n(\mu_1,\mu_2)$ are collected in Appendix~\ref{app:RGEstuff}. We use (\ref{Umatr}) to perform the evolution of the Wilson coefficients in the intervals between flavor thresholds. Concretely, we calculate the initial conditions for the coefficient functions at a high scale $\mu_Z\sim m_Z$ from (\ref{Cndef}). Notice that these relations are free of large logarithms at this scale. We then evolve the coefficients to a matching scale $\mu_b\sim m_b$ using (\ref{Umatr}) and evaluating the evolution functions with $n_f=5$ active quark flavors. At the scale $\mu_b$ the $b$ quark is integrated out from the effective theory, and in this process we need to perform a careful matching of the coefficients in the 5-flavor and 4-flavor theories. The relevant matching conditions are 
\begin{equation}
\begin{aligned}
   C_n^S(\mu_b) \big|_{n_f=4}
   &= \frac{{\cal Q}_S^{(5)}}{{\cal Q}_S^{(4)}}\,C_n^S(\mu_b) \big|_{n_f=5}
    + \bigg( 1 - \frac{{\cal Q}_S^{(5)}}{{\cal Q}_S^{(4)}} \bigg) C_n(\mu_b) \big|_{n_f=5} \,, \\
   D_n(\mu_b) \big|_{n_f=4}
   &= \frac{{\cal Q}_S^{(5)}}{{\cal Q}_S^{(4)}}\,D_n(\mu_b) \big|_{n_f=5} \,, \qquad
    C_n(\mu_b) \big|_{n_f=4} = C_n(\mu_b) \big|_{n_f=5} \,, \\
   C_n^b(\mu_b) &= C_n(\mu_b) \big|_{n_f=5} + \frac{{\cal Q}_S^{(5)}}{Q_b v_b}
    \left[ C_n^S(\mu_b) - C_n(\mu_b) \right] \big|_{n_f=5} \,.
\end{aligned}
\end{equation}
We then continue to run the coefficients down to a matching scale $\mu_c$ using (\ref{Umatr}) with evolution functions corresponding to $n_f=4$ active quark flavors. At the scale $\mu_c$ the charm quark is integrated out. The relevant matching conditions are
\begin{equation}
\begin{aligned}
   C_n^S(\mu_c) \big|_{n_f=3}
   &= \frac{{\cal Q}_S^{(4)}}{{\cal Q}_S^{(3)}}\,C_n^S(\mu_c) \big|_{n_f=4}
    + \bigg( 1 - \frac{{\cal Q}_S^{(4)}}{{\cal Q}_S^{(3)}} \bigg) C_n(\mu_c) \big|_{n_f=4} \,, \\
   D_n(\mu_c) \big|_{n_f=3}
   &= \frac{{\cal Q}_S^{(4)}}{{\cal Q}_S^{(3)}}\,D_n(\mu_c) \big|_{n_f=4} \,, \qquad
    C_n(\mu_c) \big|_{n_f=3} = C_n(\mu_c) \big|_{n_f=4} \,, \\
   C_n^c(\mu_c) &= C_n(\mu_c) \big|_{n_f=4} + \frac{{\cal Q}_S^{(4)}}{Q_c v_c}
    \left[ C_n^S(\mu_c) - C_n(\mu_c) \right] \big|_{n_f=4} \,.
\end{aligned}
\end{equation}
Finally, we run the coefficients down to the low-energy scale $\mu_0$ using (\ref{Umatr}) with evolution functions corresponding to $n_f=3$ active quark flavors. The values we obtain for the Wilson coefficients of the first few Gegenbauer moments at the scale $\mu_0=1$~GeV are collected in Table~\ref{tab:Cnvals}. Two points are worth mentioning. First, for $n\ge 2$ the coefficients $D_n$ of the gluon LCDA are of the same magnitude as those of the quark LCDAs, which is remarkable given that the former ones start at ${\cal O}(\alpha_s)$, whereas the latter ones start at tree level. Second, we observe that the differences between the flavor-singlet coefficients $C_n^S$ and the flavor-nonsinglet coefficients $C_n$ are numerically very small. This shows that the mixing of quark and gluon LCDAs under RG evolution is a small effect. 

\begin{table}
\centering
\begin{tabular}{|c|cccc|}
\hline 
$n$ & 0 & 2 & 4 & 6 \\
\hline
$C_n^S(\mu_0)$ & $0.937$ & $0.413+0.063i$ & $0.291+0.061i$ & $0.233+0.055i$ \\ 
% $C_n^{\rm NLO}(\mu_0)$ & $0.837$ & & & \\
$C_n(\mu_0)$ & $0.937$ & $0.409+0.064i$ & $0.290+0.061i$ & $0.232+0.055i$ \\
% $C_n^{\rm NLO}(\mu_0)$ & $0.937$ & $0.379+0.064i$ & $0.266+0.061i$ & $0.212+0.055i$ \\
$C_n^b(\mu_b)$ & $0.937$ & $0.658+0.101i$ & $0.579+0.121i$ & $0.535+0.127i$ \\
$C_n^c(\mu_c)$ & $0.937$ & $0.464+0.071i$ & $0.346+0.072i$ & $0.287+0.068i$ \\
$D_n(\mu_0)$ & $0$ & $0.430+0.024i$ & $0.265+0.043i$ & $0.192+0.040i$ \\ 
\hline
\end{tabular}
\parbox{15.5cm}
{\caption{\label{tab:Cnvals} 
Short-distance coefficients entering the factorization formula (\ref{Fetanf3}) evaluated at the low scale $\mu_0=1$~GeV. We use $\mu_Z=m_Z$, as well as $\mu_b=m_b=4.163$~GeV and $\mu_c=m_c=1.279$~GeV for the heavy-flavor thresholds.}}
\end{table} 

Beyond one-loop order the solution of the RG equation (\ref{RGevol1n}) takes a much more complicated form, since Gegenbauer moments of different rank mix under renormalization. The corresponding expressions for the singlet and nonsinglet cases can be derived from \cite{Mertig:1995ny,Belitsky:1998uk} and are collected in the appendix of \cite{Agaev:2014wna}. A dedicated study of NLO evolution effects for the case of exclusive radiative Higgs-boson decays has been performed in \cite{Koenig:2015pha}. It turns out these effects have only a modest impact on our results. For example, the values of the nonsinglet coefficients $C_n(\mu_0)$ shown in Table~\ref{tab:Cnvals} are reduced by factors of 1, 0.93, 0.92, 0.91 for $n=0,2,4,6$ when NLO evolution effects are taken into account. The most important impact on the singlet coefficients is due to the anomalous dimension of the flavor-singlet axial current, which starts at two-loop order \cite{Kodaira:1979pa}. As a result, when NLO evolution effects are included the value of the leading singlet coefficient $C_0^S(\mu_0)$ differs from the value given in Table~\ref{tab:Cnvals} by the factor 
\begin{equation}
   \kappa_{\rm NLO}
   = 1 + \frac{30}{23}\,\frac{\alpha_s(\mu_Z)}{\pi} - \frac{198}{575}\,\frac{\alpha_s(\mu_b)}{\pi}
    - \frac{22}{75}\,\frac{\alpha_s(\mu_c)}{\pi} - \frac{2}{3}\,\frac{\alpha_s(\mu_0)}{\pi}
   \approx 0.89 \,.
\end{equation}
Given the present large uncertainties in the values of the Gegenbauer moments with $n\ne 0$, it is a reasonable approximation to account for the effects of NLO evolution by treating $\kappa_{\rm NLO}$ as a global prefactor in the expressions for the form factors and otherwise use the coefficients compiled in Table~\ref{tab:Cnvals}. The deviation of $\kappa_{\rm NLO}$ from~1 provides a measure of the remaining perturbative uncertainty in our predictions.

\subsection{Hadronic input parameters}
\label{sec:mesonMixing}

It remains to obtain the required hadronic input for the decay constants and Gegenbauer moments of the $\eta$ and $\eta'$ mesons. Following \cite{Beneke:2002jn}, we assume isospin symmetry of all hadronic matrix elements, but we differentiate between the matrix elements of operators containing up and down quarks and those containing strange quarks. In the $SU(3)$ flavor-symmetry limit, the pseudoscalar meson $\eta$ would be a flavor octet and $\eta'$ a flavor singlet. However, it is known empirically that $SU(3)$-breaking corrections to these assignments are large. In the following we shall thus not rely on $SU(3)$ flavor symmetry but instead introduce another assumption, expected to be accurate at the 10\% level. In the absence of the axial $U(1)$ anomaly, the flavor states $|\eta_q\rangle=(|u\bar u\rangle+|d\bar d\rangle)/\sqrt2$ and $|\eta_s\rangle=|s\bar s\rangle$ mix only through OZI-violating effects, which are known phenomenologically to be small. It is therefore reasonable to assume that the axial anomaly is the only effect that mixes these two flavor states. This is the basis of the FKS mixing scheme \cite{Feldmann:1998vh}. Since this is by assumption the only mixing effect, the FKS scheme amounts to a scheme with a single mixing angle, which relates the physical mass eigenstates to the flavor states via
\begin{equation}\label{etaetap}
   \left( \begin{array}{c} |\eta\rangle \\ |\eta'\rangle \end{array} \right)
   = \left( \begin{array}{lr} \cos\varphi & ~-\sin\varphi \\ \sin\varphi & \cos\varphi \end{array} \right)
    \left( \begin{array}{c} |\eta_q\rangle \\ |\eta_s\rangle \end{array} \right) .
\end{equation}

In the FKS mixing scheme, one introduces decay constants $f_q$ and $f_s$ and light-cone distribution amplitudes $\phi_q(x,\mu)$ and $\phi_s(x,\mu)$ for the flavor states $|\eta_{q,s}\rangle$ \cite{Beneke:2002jn}. The physical $\eta$ and $\eta'$ mesons are then described as coherent superpositions of these states. It follows that the flavor-specific decay constants defined in (\ref{fPqdef}) are given by
\begin{equation}
\begin{aligned}
   f_\eta^u = f_\eta^d &= \frac{f_q}{\sqrt2}\,\cos\varphi \,, & \quad
    f_\eta^s &= - f_s \sin\varphi \,, \\[-1mm]
   f_{\eta'}^u = f_{\eta'}^d &= \frac{f_q}{\sqrt2}\,\sin\varphi \,, &
    f_{\eta'}^s &= f_s \cos\varphi \,.
\end{aligned}
\end{equation}
Analogous relations hold for the flavor-specific LCDAs defined in (\ref{quarkLCDA}), i.e.\
\begin{equation}
\begin{aligned}
   f_\eta^u\,\phi_\eta^u = f_\eta^d\,\phi_\eta^d 
   &= \frac{f_q\,\phi_q}{\sqrt2}\,\cos\varphi \,, & \quad
    f_\eta^s\,\phi_\eta^s &= - f_s\,\phi_s \sin\varphi \,, \\[-1mm]
   f_{\eta'}^u\,\phi_{\eta'}^u = f_{\eta'}^d\phi_{\eta'}^d 
   &= \frac{f_q\,\phi_q}{\sqrt2}\,\sin\varphi \,, &
    f_{\eta'}^s\,\phi_{\eta'}^s &= f_s\,\phi_s \cos\varphi \,.
\end{aligned}
\end{equation}
Moreover, in the FKS scheme one sets $\phi_\eta^g(x,\mu)=\phi_{\eta}^g(x,\mu)\equiv\phi_g(x,\mu)$ for the gluon LCDAs \cite{Beneke:2002jn,Kroll:2013iwa,Agaev:2014wna}. As a consequence, the hadronic parameters characterizing the quark and gluon LCDAs are the Gegenbauer moments $a_n^q$ and $a_n^s$ of $\phi_q$ and $\phi_s$ and the Gegenbauer moments $b_n$ of $\phi_g$, defined as in (\ref{eq:LCDAexpansion}).

The decay constants $f_q$, $f_s$ and the mixing angle $\varphi$ in the FKS scheme have been determined in \cite{Feldmann:1998vh} from a fit to experimental data, finding
\begin{equation}
   f_q = (1.07\pm 0.02)\,f_\pi\,, \qquad
   f_s = (1.34\pm 0.06)\,f_\pi \,, \qquad
   \varphi = 39.3^\circ\pm 1.0^\circ \,.\label{FKS1}
\end{equation}
Here $f_\pi=(130.4\pm 0.2)$~MeV is the pion decay constant \cite{Agashe:2014kda}. A more recent analysis exploiting additional data but only a subset of the processes investigated in the original paper finds \cite{Escribano:2005qq}
\begin{equation}
   f_q = (1.09\pm 0.03)\,f_\pi\,, \qquad
   f_s = (1.66\pm 0.06)\,f_\pi \,, \qquad
   \varphi = 40.7^\circ\pm 1.4^\circ \,.\label{FKS2}
\end{equation}
The central values of the flavor-singlet decay constants $f_{\eta^{(\prime)}}^{uds}$, which provide the normalization for the leading contributions to the form factors in (\ref{Fetanf3}), are $f_\eta^{uds}=42.0$~MeV and $f_{\eta'}^{uds}=260.2$~MeV for the parameters in (\ref{FKS1}), and $f_\eta^{uds}=11.2$~MeV and $f_{\eta'}^{uds}=295.2$~MeV when employing the parameters in (\ref{FKS2}). The drastic sensitivity of $f_\eta^{uds}$ to the choice of FKS parameters will be reflected in the spread of our predictions for the $Z\to\eta\gamma$ branching ratio. 

The Gegenbauer moments of the LCDAs can be extracted using fits to data for the $\gamma^*\gamma\to\eta^{(\prime)}$ transition form factors at different $Q^2$ reported by the CLEO \cite{Gronberg:1997fj} and BaBar \cite{BABAR:2011ad} collaborations. The authors of \cite{Agaev:2014wna} have assumed $SU(3)$ flavor symmetry and have chosen the first few Gegenbauer moments of the quark LCDAs $\phi_q$ and $\phi_s$ according to some popular QCD sum-rule calculations for the pion LCDA. The first Gegenbauer moment of the gluon LCDA $\phi_g$ has then been extracted from the fit to the data. In this context three benchmark models were identified, to which we refer below as models (i)--(iii). In all three cases the FKS parameters in (\ref{FKS1}) have been used. The authors of \cite{Kroll:2013iwa}, on the contrary, have extracted the first Gegenbauer moments of both the quark and the gluon LCDAs from fits to the data. Below we will consider three of their fit scenarios. Model (iv) refers to their default fit, while model (v) corresponds to a fit exclusively to BaBar data. In both cases the FKS parameters in (\ref{FKS1}) are assumed. Model (vi) corresponds to a combined fit to CLEO and BaBar data using the FKS parameters in (\ref{FKS2}). Table~\ref{tab:gegenbauer} collects the values of the Gegenbauer moments in the six benchmark models, translated to our notations.

\begin{table}
\centering
\begin{tabular}{|c|ccccc|c|}
\hline
 Model& $a_2^q$ & $a_4^q$ & $a_2^s$ & $a_4^s$ & $b_2$ & FKS pars.\ \\ 
\hline
(i) & 0.10 & 0.10 & 0.10 & 0.10 & $-0.06$ & (\ref{FKS1}) \\
(ii) & 0.20 & 0.00 & 0.20 & 0.00 & $-0.07$ & (\ref{FKS1}) \\
(iii) & 0.25 & $\hspace{-3mm}-0.10$ & 0.25 & $\hspace{-3mm}-0.10$ & $-0.06$ & (\ref{FKS1}) \\ 
\hline
(iv) & $-0.10$ & & $-0.07$ & & $-0.14$ & (\ref{FKS1}) \\
(v) & $-0.10$ & & $-0.07$ & & $-0.24$ & (\ref{FKS1}) \\ 
(vi) & $-0.09$ & & $-0.02$ & & $-0.08$ & (\ref{FKS2}) \\ 
\hline
\end{tabular} 
\parbox{15.5cm}
{\caption{\label{tab:gegenbauer} 
Gegenbauer moments of quark and gluon LCDAs at the scale $\mu_0=1$~GeV in different benchmark models obtained from analyses of the $\gamma^*\gamma\to\eta^{(\prime)}$ transition form factors. Models (i)--(iii) correspond to the models in Table~2 of \cite{Agaev:2014wna}, while models (iv)--(vi) refer to the first, third and sixth model in Table~2 of~\cite{Kroll:2013iwa}.}}
\end{table}

To evaluate the form factors in (\ref{Fetanf3}) we also need the decay constants $f_P^c$ and $f_P^b$ describing the intrinsic charm and bottom contents of the $\eta$ and $\eta'$ mesons. Following \cite{Beneke:2002jn}, we estimate these parameters using relations among the FKS parameters implied by the axial anomaly. This yields
\begin{equation}
   f_P^c(\mu_c)\approx - \frac{m_P^2}{12m_c^2}\,f_P^u \,, \qquad
   f_P^b(\mu_b)\approx - \frac{m_P^2}{12m_b^2}\,f_P^u \,.
\end{equation}
Numerically, we obtain $f_\eta^c\approx-1.2$~MeV, $f_{\eta'}^c\approx-2.9$~MeV and $f_\eta^b\approx-0.1$~MeV, $f_{\eta'}^b\approx-0.3$~MeV. These values are of the same order as those obtained using similar methods in \cite{Feldmann:1998vh,Yuan:1997ts,Ali:1997ex,Petrov:1997yf,Franz:2000ee}. Due to the very small effect of the intrinsic charm and bottom contributions on the form factors we only keep the leading terms with $n=0$ in these contributions and set all Gegenbauer moments $a_n^{P,c}$ and $a_n^{P,b}$ with $n>0$ to zero.

\section{Phenomenological predictions}

We are now ready to present our numerical results. Before we quote our predictions for the $Z\to\eta^{(\prime)}\gamma$ branching ratios, we demonstrate the sensitivity of the form factors in (\ref{Fetanf3}) to the choice of the FKS parameters, the decay constants $f_P^{c,b}$, the Gegenbauer moments of the LCDAs and the factorization scale. Using the mixing parameters in (\ref{FKS1}) as our default values, we obtain for the real parts of the form factors
\begin{equation}\label{eq:FormFKS1}
\begin{aligned}
   \frac{\mbox{Re}\,{\cal F}_\eta}{\kappa_{\rm NLO}} 
   &= 16.3\,\mbox{MeV}\,\Big( 1 + 1.41 a_2^q - 0.97 a_2^s + 0.40 b_2 + 0.99 a_4^q - 0.68 a_4^s
    + 0.25 b_4 + \dots \Big) + \delta_\eta \,, \\
   \frac{\mbox{Re}\,{\cal F}_{\eta'}}{\kappa_{\rm NLO}}
   &= 86.5\,\mbox{MeV}\,\Big( 1 + 0.22 a_2^q + 0.22 a_2^s + 0.46 b_2 + 0.15 a_4^q + 0.16 a_4^s
    + 0.29 b_4 + \dots \Big) + \delta_{\eta'} .
\end{aligned}
\end{equation}
The small imaginary parts of ${\cal O}(\alpha_s)$ can be neglected, since they do not contribute to the decay rates at NLO in RG-improved perturbation theory. The parameters $f_q$, $f_s$ and $\varphi$ used to compute these values come with uncertainties, which we did not take into account in the expressions above for readability. We demonstrate their effect by looking only at the leading terms, for which we find
\begin{equation}\label{eq40}
\begin{aligned}
   \mbox{Re}\,{\cal F}_\eta 
   &= \kappa_{\rm NLO} \left( 16.3\pm 1.5_\varphi\pm 1.0_{f_q}\pm 1.6_{f_s} \right) \mbox{MeV} + \dots \,, \\
   \mbox{Re}\,{\cal F}_{\eta'} 
   &= \kappa_{\rm NLO} \left( 86.5\pm 0.3_\varphi\pm 0.8_{f_q}\pm 2.0_{f_s} \right) \mbox{MeV} + \dots \,.
\end{aligned}
\end{equation}
We now recompute these results using the FKS parameters collected in (\ref{FKS2}). In this case we find
\begin{equation}\label{eq:FormFKS2}
\begin{aligned}
   \frac{\mbox{Re}\,{\cal F}_\eta}{\kappa_{\rm NLO}} 
   &= 6.3\,\mbox{MeV}\,\Big( 1 + 3.64 a_2^q - 3.20 a_2^s + 0.27 b_2 + 2.56 a_4^q - 2.25 a_4^s
    + 0.17 b_4 + \dots \Big) + \delta_\eta \,, \\
   \frac{\mbox{Re}\,{\cal F}_{\eta'}}{\kappa_{\rm NLO}} 
   &= 98.0\,\mbox{MeV}\,\Big( 1 + 0.20 a_2^q + 0.24 a_2^s + 0.46 b_2 + 0.14 a_4^q + 0.17 a_4^s
    + 0.29 b_4 + \dots \Big) + \delta_{\eta'} ,
\end{aligned} 
\end{equation}
and the parametric uncertainties in the leading terms read
\begin{equation}\label{eq42}
\begin{aligned}
   \mbox{Re}\,{\cal F}_\eta 
   &= \kappa_{\rm NLO} \left( 6.3\pm 2.4_\varphi\pm 1.4_{f_q}\pm 1.7_{f_s} \right) \mbox{MeV} +\dots \,, \\
   \mbox{Re}\,{\cal F}_{\eta'} 
   &= \kappa_{\rm NLO} \left( 98.0\,{}^{+0.1}_{-0.2}{}_{\,\varphi}\pm 1.2_{f_q}\pm 1.9_{f_s}\right) \mbox{MeV} + \dots \,. 
\end{aligned}
\end{equation}
A comparison of relations (\ref{eq:FormFKS1}) with (\ref{eq:FormFKS2}), and (\ref{eq40}) with (\ref{eq42}) reveals a very strong sensitivity of the form factor ${\cal F}_\eta$ to the employed mixing parameters. The results for ${\cal F}_{\eta'}$ are more stable, but the difference between the central values shown in (\ref{eq40}) and (\ref{eq42}) is still much larger than the parametric uncertainties within each parameter set. Since the dominant contributions to the form factors are proportional to the flavor-singlet decay constants $f_P^{uds}$, the large variations of $f_P^{uds}$ observed earlier translate directly into variations of the form factors. A future precise measurement of the $Z\to\eta\gamma$ branching ratio could thus give important insights on the correct values of the FKS parameters. In the above expressions we have quoted the dependence of the form factors on the intrinsic charm and bottom content separately. They are contained 
in the parameters $\delta_P\approx 0.36 f^c_P+0.32 f^b_P$, for which we obtain $\delta_\eta\approx-0.45$~MeV and $\delta_{\eta'}\approx-1.2$~MeV. These effects are rather small numerically. 

The contributions from the gluon LCDA to the form factors are important. For the $\eta'$ meson the coefficients of the gluon Gegenbauer moments $b_2$ and $b_4$ are significantly larger than those of the quark Gegenbauer moments of the same order. In the case of the $\eta$ meson the individual quark Gegenbauer moments come with large coefficients, but there are large cancellations between the terms involving $a_n^q$ and $a_n^s$ with the same $n$. If one assumes $SU(3)$ flavor symmetry as in \cite{Agaev:2014wna}, then the contributions of the quark and gluon Gegenbauer moments at the same order come again with similar coefficients, but obviously $SU(3)$-breaking corrections could have a large impact. If future theoretical efforts based on lattice gauge theory or refined QCD sum rules can provide a better control of the Gegenbauer moments of the quark LCDAs, then both decay channels could potentially provide valuable information about the moments of the gluon LCDA.

Compared with the significant hadronic uncertainties we have encountered, the perturbative uncertainties of the form factors as estimated from scale variations are very small. We illustrate this by showing the scale uncertainties of the dominant terms in (\ref{eq:FormFKS1}) obtained by varying the factorization scale $\mu_Z$ between $m_Z/2$ and $2m_Z$. We find
\begin{equation}
\begin{aligned}
   \frac{\mbox{Re}\,{\cal F}_\eta}{\kappa_{\rm NLO}} 
   &= 16.3\,\mbox{MeV}\,\Big[ (1\pm0.01) + \left( 1.41^{+0.01}_{-0.02} \right) a_2^q
    - \left( 0.97\pm 0.01 \right) a_2^s + \left( 0.40^{+0.00}_{-0.02} \right) b_2 + \dots \Big] \,, \\
   \frac{\mbox{Re}\,{\cal F}_{\eta'}}{\kappa_{\rm NLO}}
   &= 86.5\,\mbox{MeV}\,\Big[ (1\pm0.01) + \left(0.22^{+0.00}_{-0.01} \right) a_2^q
    - \left( 0.22\pm0.00 \right) a_2^s + \left( 0.46^{+0.00}_{-0.02} \right) b_2 + \dots \Big] \,.
\end{aligned}
\end{equation}
In light of the large parametric uncertainties and the estimated size of NLO evolution effects, we find that the scale uncertainties have a negligible impact on our predictions.

Given our results for the form factors ${\cal F}_P$, the $Z\to P\gamma$ branching ratios are obtained as\begin{equation}
   \mathrm{Br}(Z\to P\gamma) = \frac{\alpha m_Z}{6 v^2\Gamma_Z} \left| {\cal F}_P \right|^2 ,
\end{equation}
where $\alpha=1/137.036$ is the electromagnetic fine structure constant evaluated at $q^2=0$, $\Gamma_Z=2.4955$~GeV is the $Z$-boson width, and $v=245.36$~GeV is the Higgs vacuum expectation value \cite{Agashe:2014kda}. Table~\ref{tab:branching} shows our predictions for the $Z\to\eta^{(\prime)}\gamma$ branching fractions obtained using the hadronic parameters corresponding to the six models collected in Table~\ref{tab:gegenbauer} along with $\kappa_{\rm NLO}=0.89$. The sensitivity of the branching ratios to the hadronic input parameters opens up the possibility of probing these parameters using the decays $Z\to\eta^{(\prime)}\gamma$. This is particularly interesting since the different LCDA parameters from Table~\ref{tab:gegenbauer} are all obtained from fits to the same low-energy experimental data, but still differ a lot from each other depending on the assumptions and methods used to perform these fits. Owing to the very large value $\mu_Z\sim m_Z$ of the factorization scale inherent to $Z\to P\gamma$ decays, the analysis of these processes is much cleaner theoretically and less affected by uncertainties due to power corrections or higher-order perturbative corrections.

\begin{table}
\centering
\begin{tabular}{|c|ccc|ccc|}
\hline
Model & (i) & (ii) & (iii) & (iv) & (v) & (vi) \\
\hline
$\hspace{-1mm}\mathrm{Br}(Z\to\eta\gamma)\hspace{-1mm}$
 & $0.16\pm 0.05$ & $0.17\pm 0.05$ & $0.16\pm 0.05$
 & $0.11\pm 0.03$ & $0.10\pm 0.03$ & $0.010\,_{-0.010}^{+0.014}$ \\
\hline
$\hspace{-1mm}\mathrm{Br}(Z\to\eta^\prime\gamma)\hspace{-1mm}$ 
 & $4.70\pm 0.23$ & $4.77\pm 0.24$ & $4.73\pm 0.24$
 & $3.43\pm 0.17$ & $3.08\pm 0.15$ & $4.84\pm 0.23$ \\ 
\hline
\end{tabular}
\parbox{15.5cm}
{\caption{\label{tab:branching} 
Central values of the $Z\to\eta^{(\prime)}\gamma$ branching ratios in units of $10^{-9}$, obtained using six different models of hadronic input parameters, see Table~\ref{tab:gegenbauer}. Models (i)--(v) use the mixing parameters in (\ref{FKS1}), while model (vi) uses those in (\ref{FKS2}). We take $\kappa_{\rm NLO}=0.89$ to account for NLO evolution effects.}}
\end{table} 

The branching ratios we obtain are of similar magnitude as those for the corresponding decays to light vector mesons found in \cite{Grossmann:2015lea}, e.g.\ $\mbox{Br}(Z\to\rho\gamma)\approx 4.2\cdot 10^{-9}$ and $\mbox{Br}(Z\to\phi\gamma)\approx 8.6\cdot 10^{-9}$. A future high-luminosity $e^+ e^-$ collider operating at the $Z$ pole could produce samples of about $10^{12}$ $Z$ bosons per year \cite{Blondel:2013rn}. This would yield at best ${\cal O}(150)$ events in the $Z\to\eta\gamma$ channel and ${\cal O}(4500)$ events in 
the $Z\to\eta'\gamma$ channel without considering backgrounds and reconstruction efficiencies.

\section{Conclusions}

We have presented a detailed analysis of the rare radiative decays $Z\to\eta\gamma$ and $Z\to\eta'\gamma$ within the QCD factorization approach, working at NLO in RG-improved perturbation theory. In particular, we have included the additional contributions that arise since the final-state mesons $\eta$ and $\eta'$ contain a flavor-singlet component in their wave function, which can be formed from a two-gluon state. We have derived the corresponding QCD factorization formula in the framework of SCET, generalizing the derivation presented in \cite{Grossmann:2015lea} to the flavor-singlet case. In the presence of the gluon contribution the scaling behavior of the hadronic matrix elements is changed, because quark and gluon LCDAs mix under renormalization. This complicates the resummation of large logarithms. Since the flavor singlet is composed of a different number of dynamical quark flavors at the high scale $\mu\sim m_Z$ compared to the low hadronic scale $\mu\sim 1$~GeV, the singlet and nonsinglet matrix elements need to be rearranged when crossing quark flavor thresholds, giving rise to non-trivial matching conditions. Also, the intrinsic charm and bottom content of the $\eta^{(\prime)}$ mesons gives rise to non-trivial effects.

Many of the hadronic input parameters needed for a phenomenological analysis of our calculations are not yet well determined. In the literature, these parameters have been obtained from different approaches leading to sometimes quite different results. Considering a set of six benchmark models obtained in \cite{Agaev:2014wna,Kroll:2013iwa}, the values we obtain for the $Z\to\eta\gamma$ branching ratio range from $0.1\cdot 10^{-10}$ to $1.7\cdot 10^{-10}$, while those for the $Z\to\eta'\gamma$ branching fraction vary between 
$3.1\cdot 10^{-9}$ and $4.8\cdot 10^{-9}$. We find that especially the $Z\to\eta\gamma$ channel is very sensitive to the $\eta\!-\!\eta'$ mixing parameters, so that a measurement of this decay mode could be used to favor one particular set of mixing parameters over another. A similar statement can be made with regard to the different sets of LCDA shape parameters. While all models in Table~\ref{tab:gegenbauer} seem to give consistent descriptions of the $\gamma^*\gamma\to\eta^{(\prime)}$ transition form factors, we find that some of them give rather distinct predictions for the $Z\to\eta^{(\prime)}\gamma$ branching ratios. Measurements of these rare radiative $Z$-boson decays could thus be used as complementary information to attain a better control over the hadronic matrix elements describing the $\eta\!-\!\eta'$ system. If in the future the quark LCDAs and meson mixing parameters can be determined more accurately, using more precise low-energy data and advanced theoretical tools such as lattice gauge theory, a study of the $Z\to\eta^{(\prime)}\gamma$ decay modes at a future lepton collider would provide a very good opportunity to directly access the Gegenbauer moments of the gluon LCDA in a theoretically clean environment.

\subsubsection*{Acknowledgments}
We are grateful to Martin Beneke for useful discussions. This work was supported by the Advanced Grant EFT4LHC of the European Research Council (ERC), the Cluster of Excellence {\em Precision Physics, Fundamental Interactions and Structure of Matter} (PRISMAâ EXC 1098), grant 05H12UME of the German Federal Ministry for Education and Research (BMBF) and the DFG Graduate School {\em Symmetry Breaking in Fundamental Interactions} (GRK 1581).

\begin{appendix}
\numberwithin{equation}{section}

\section{Renormalization-group evolution}
\label{app:RGEstuff}

The explicit expressions for the kernel functions defined in (\ref{eq:LCDArenZ}) are \cite{Terentev:1980qu,Ohrndorf:1981uz,Shifman:1980dk,Baier:1981pm}
\begin{equation}
\begin{aligned}
   V_{qq}(x,y) &= -2 C_F\,\bigg[ \frac{x}{y} \left( 1 + \frac{1}{y-x} \right) \theta(y-x)
    + \bigg( \begin{array}{c} x\to 1-x \\ y\to 1-y \end{array} \bigg) \bigg]_+ \,, \\
   V_{qg}(x,y) &= 8 T_F n_f\,\bigg[ \frac{x}{y^2}\,\theta(y-x)
    - \bigg( \begin{array}{c} x\to 1-x \\ y\to 1-y \end{array} \bigg) \bigg] \,, \\
   V_{gq}(x,y) &= - C_F\,\bigg[ \frac{x^2}{y}\,\theta(y-x)
    - \bigg( \begin{array}{c} x\to 1-x \\ y\to 1-y \end{array} \bigg) \bigg] \,, \\
   V_{gg}(x,y) &= - \beta_0\,\delta(x-y) - 2C_A\,\bigg\{ 
    \frac{x}{y} \left[ \left( \frac{\theta(y-x)}{y-x} \right)_+ +\frac{2x-1}{y}\,\theta(y-x) 
    \right] + \bigg( \begin{array}{c} x\to 1-x \\ y\to 1-y \end{array} \bigg) \bigg\} \,,
\end{aligned}
\end{equation}
where as usual the plus distribution is defined as
\begin{equation}
   \big[ F(x,y) \big]_+ = F(x,y) - \delta(x-y) \int_0^1\!dz\,F(z,y) \,.
\end{equation}
At leading order in RG-improved perturbation theory, the explicit solution for the evolution matrix $\bm{U}_n^S(\mu_1,\mu_2)$ in (\ref{Umatr}) is given by \cite{Terentev:1980qu,Ohrndorf:1981uz,Shifman:1980dk,Baier:1981pm}
\begin{equation}
\begin{aligned}
   \bm{U}_n^S(\mu_1,\mu_2)
   &= \exp\bigg[ - \frac{1}{2\beta_0}\,\ln\frac{\alpha_s(\mu_1)}{\alpha_s(\mu_2)}
     \begin{pmatrix} \gamma_n^{qq} & ~\gamma_n^{qg} \\
                     \gamma_n^{gq} & ~\gamma_n^{gg}
     \end{pmatrix}^T \bigg] \\ 
   &= \begin{pmatrix} \frac{1+r_n}{2}
    \Big( \frac{\alpha_s(\mu_2)}{\alpha_s(\mu_1)} \Big)^{\frac{\gamma_n^+}{2\beta_0}}
    + \frac{1-r_n}{2} \Big( \frac{\alpha_s(\mu_2)}{\alpha_s(\mu_1)} \Big)^{\frac{\gamma_n^-}{2\beta_0}}
    &\quad& \frac{\gamma_n^{gq}}{\gamma_n^+ -\gamma_n^-} \bigg[
     \Big( \frac{\alpha_s(\mu_2)}{\alpha_s(\mu_1)} \Big)^{\frac{\gamma_n^+}{2\beta_0}}
     - \Big( \frac{\alpha_s(\mu_2)}{\alpha_s(\mu_1)} \Big)^{\frac{\gamma_n^-}{2\beta_0}} \bigg] \\[4mm]
    \frac{\gamma_n^{qg}}{\gamma_n^+ -\gamma_n^-} \bigg[
     \Big( \frac{\alpha_s(\mu_2)}{\alpha_s(\mu_1)} \Big)^{\frac{\gamma_n^+}{2\beta_0}}
     - \Big( \frac{\alpha_s(\mu_2)}{\alpha_s(\mu_1)} \Big)^{\frac{\gamma_n^-}{2\beta_0}} \bigg]
    &\quad& \frac{1-r_n}{2} \Big( \frac{\alpha_s(\mu_2)}{\alpha_s(\mu_1)} \Big)^{\frac{\gamma_n^+}{2\beta_0}}
     + \frac{1+r_n}{2} \Big( \frac{\alpha_s(\mu_2)}{\alpha_s(\mu_1)} \Big)^{\frac{\gamma_n^-}{2\beta_0}}
    \end{pmatrix} ,
\end{aligned}   
\end{equation}
where
\begin{equation}
   r_n = \frac{\gamma_n^{qq}-\gamma_n^{gg}}{\gamma_n^+ - \gamma_n^-} \,, \qquad   
   \gamma_n^\pm = \frac12 \left( \gamma_n^{qq} + \gamma_n^{gg} 
    \pm \sqrt{\big( \gamma_n^{qq} - \gamma_n^{gg} \big)^2 + 4\gamma_n^{qg}\gamma_n^{gq}} \right) .
\end{equation}
Here $\gamma_n^\pm$ are the eigenvalues of the one-loop anomalous-dimension matrix. The corresponding solution of the evolution function $U_n(\mu_1,\mu_2)$ is much simpler and reads
\begin{equation}
   U_n(\mu_1,\mu_2)
   = \exp\bigg[ - \frac{\gamma_n^{qq}}{2\beta_0}\,\ln\frac{\alpha_s(\mu_1)}{\alpha_s(\mu_2)} \bigg] 
   = \bigg( \frac{\alpha_s(\mu_2)}{\alpha_s(\mu_1)} \bigg)^{\frac{\gamma_n^{qq}}{2\beta_0}} \,.
\end{equation}

\end{appendix}

\end{document}